\documentclass[pra,twocolumn,showpacs,preprintnumbers,superscriptaddress,nofootnoteinbib]{revtex4}
\usepackage{bm}
\usepackage{graphicx}
\usepackage{amsbsy}
\usepackage{amsmath}
\usepackage{amsfonts}
\usepackage{amsthm}

\begin{document}

\title{Entanglement of Assistance is not a
bipartite measure nor a tripartite monotone}
\author{Gilad Gour}
\email{ggour@math.ucsd.edu}
\affiliation{Department of Mathematics,
University of California/San Diego, La Jolla, California
92093-0112}
\author{Robert W. Spekkens}
\email{rspekkens@perimeterinstitute.ca} \affiliation{Perimeter
Institute for Theoretical Physics, Waterloo, Ontario N2L 2Y5,
Canada}\affiliation{Institute for Quantum Computing, University of
Waterloo, Ontario N2L 3G1, Canada}
\date{Dec. 13, 2005}

\begin{abstract}
The entanglement of assistance quantifies the entanglement that
can be generated between two parties, Alice and Bob, given
assistance from a third party, Charlie, when the three share a
tripartite state and where the assistance consists of Charlie
initially performing a measurement on his share and communicating
the result to Alice and Bob through a one-way classical channel.
We argue that if this quantity is to be considered an operational
measure of entanglement, then it must be understood to be a
\emph{tripartite} rather than a \emph{bipartite} measure. We
compare it with a distinct tripartite measure that quantifies the
entanglement that can be generated between Alice and Bob when they
are allowed to make use of a two-way classical channel with
Charlie. We show that the latter quantity, which we call the
\emph{entanglement of collaboration}, can be greater than the
entanglement of assistance. This demonstrates that the
entanglement of assistance (considered as a tripartite measure of
entanglement), and its multipartite generalizations such as the
localizable entanglement, are not entanglement monotones, thereby
undermining their operational significance.

\end{abstract}

 \maketitle

\theoremstyle{plain} \newtheorem{theorem}{Theorem}
\newtheorem{lemma}[theorem]{Lemma}
\newtheorem{corollary}[theorem]{Corollary}
\newtheorem{conjecture}[theorem]{Conjecture}

\theoremstyle{definition} \newtheorem{definition}{Definition}

\theoremstyle{remark} \newtheorem*{remark}{Remark}
\newtheorem{example}{Example}


The first significant progress to be made on the problem of
quantifying entanglement came from the recognition that an
entangled state is a resource for distributed quantum information
processing~(QIP) using local operations and classical
communication~(LOCC)~\cite{BBPS}. This led to the idea that one
could define many operational measures of the entanglement of a
state as follows. For every QIP task imaginable, quantify the
success with which one could achieve this task given LOCC and one
or more copies of the entangled state in question. In order for
the degree of success to be uniquely defined, one must perform an
optimization over all protocols for achieving the task subject to
the LOCC restriction. Clearly then, the degree of success must be
nonincreasing under LOCC, so that \emph{if} we take a measure of
entanglement to be simply a measure of the success with which one
can achieve a distributed QIP task under LOCC, then any such
measure must be a monotone under LOCC~\cite{PR97,VP98,Vid00}.

The introduction of the notion of an entanglement monotone has
provided a framework in which to make sense of the myriad
different measures of entanglement that have been proposed in the
past few years. Because the requirement of monotonicity is
relatively weak it has tended to lead to an inclusive attitude
towards proposed measures of entanglement. It is remarkable
therefore that this very weak requirement fails to be met by an
entanglement measure that has been widely discussed, namely, the
\emph{entanglement of assistance} (EoA)~\cite{DiV98,Coh98},
as we will show. Because the EoA is a special case of the
\emph{localizable entanglement} (LE)~\cite{Pop05}, another measure
that has recently received significant attention, the latter is
also not an entanglement monotone in general.
This result calls into question the significance of the EoA and LE
as \emph{operational} measures of entanglement, that is, as
measures of a resource for distributed QIP.

The original definition of the EoA is as follows. Suppose that
Alice and Bob possess a bipartite system in the mixed state $\rho
_{AB}$ and Charlie holds a purification of this state, which is to
say that the tripartite system held by Alice, Bob and Charlie is
in a pure state $\left\vert \Psi \right\rangle _{ABC}$ such that
$\rho _{AB}=\mathrm{Tr} _{C}\left\vert \Psi \right\rangle
_{ABC}\left\langle \Psi \right\vert .$ The EoA of
Ref.~\cite{DiV98} quantifies the maximum average amount of
pure-state entanglement (quantified by the entropy of entanglement
$E_{\text{Ent}}$) that Alice and Bob can extract from $\rho _{AB}$
by Charlie performing a measurement on his system and reporting
the outcome to Alice and Bob. It follows from the
Hughston-Jozsa-Wootters~(HJW) theorem~\cite{Hug93} that
\begin{equation}
E_{\text{Ast}}(\rho _{AB})=\max_{\{p_{i},\left\vert \psi
_{i}\right\rangle _{AB}\}}\sum_{i}p_{i}E_{\text{Ent}}(\left\vert
\psi _{i}\right\rangle _{AB}) \label{originalEoA}
\end{equation}
where the entropy of entanglement of $\left\vert \psi
_{i}\right\rangle _{AB} $ is the von Neumann entropy of the
reduced density operator on $A;$ that is,
$E_{\text{Ent}}(\left\vert \psi _{i}\right\rangle _{AB})=S(\rho
_{A,i}) $ with $\rho _{A,i}\equiv \mathrm{Tr}_{B}(\left\vert \psi
_{i}\right\rangle _{AB}\left\langle \psi _{i}\right\vert )$ and
$S(\rho)\equiv -\mathrm{Tr}(\rho\log \rho).$ The maximization
in~(\ref{originalEoA}) is over all convex decompositions of $\rho
_{AB}$ into pure states; that is, all pure state ensembles
$\{p_{i},\left\vert \psi _{i}\right\rangle _{AB}\}$ such that
$\rho _{AB}=\sum_{i}p_{i}\left\vert \psi _{i}\right\rangle
_{AB}\left\langle \psi _{i}\right\vert $.

Written in this form, it might seem that the EoA is a measure of
entanglement for \emph{bipartite} states, that is, a measure of
the degree of success that can be achieved for some distributed
QIP task given $\rho _{AB}$ and LOCC.  Indeed, in the article that
introduced the EoA, it is described as "a new quantification of
entanglement of a general (mixed) quantum state for a bipartite
system"~\cite{DiV98}. However, it can happen that Alice and Bob's
bipartite system is described by the mixed state $\rho _{AB}$ even
though Charlie's system does not purify this state. For instance,
the tripartite system might be in a \emph{mixed} state $\rho
_{ABC}$ such that $\rho _{AB}=\mathrm{Tr}_{C}\rho _{ABC}.$ So we
see that the definition privded by Eq.~(\ref{originalEoA}) does
not quantify the usefulness of the bipartite state $\rho _{AB}$
but rather that of the tripartite state $\left\vert \Psi
\right\rangle_{ABC} .$ Thus, the EoA is clearly an operational
measure of entanglement for \emph{tripartite} states.

Suppose, however, that one tried to defend the notion that the EoA
is a bipartite measure of entanglement by simply taking
Eq.~(\ref{originalEoA}) as its definition, and dispensing with any
reference to a tripartite state shared with Charlie. One can do
so, but it is then \emph{trivial} to show that such a measure is
not a monotone.  It suffices to note that one can always increase
the EoA by simply discarding information.  For instance, a product
state for two qubits, which has $E_{\text{Ast}}=0$, may be
transformed into the completely mixed state, which has
$E_{\text{Ast}}=1$ (this follows from the fact that such a state
can be decomposed into an equal mixture of the four Bell states),
by simply adding noise. The problem is that the EoA is a concave
rather than a convex function~\footnote{It is interesting to
note, however, that it is not always straightforward to equate
loss of information with mixing~\cite{Ple05}.}.

It might appear that this puts the nails in the coffin of the
notion that the EoA is a bipartite measure of entanglement, but
one can in fact go much further. By adopting a very loose policy
with regard to what should be called a measure of entanglement,
one might consider dropping the requirement of monotonicity under
LOCC operations and maintaining that the EoA is a bipartite
measure of entanglement in a broader sense. However, for a measure
to be considered an operational measure of entanglement \emph{in
any sense whatsoever} it must quantify the ability of the state to
act as a nonlocal resource for \emph{some} set of restricted
operations. In other words, there must be some set of operations,
more restrictive than LOCC, such that it is nonincreasing under
these.
For instance, one might think that by forbidding the discarding of
information, to which we appealed in the example above, the EoA
might become nonincreasing under the remaining operations. As it
turns out, this is not the case.  One can find examples of states
for which the EoA increases even under LOCC with no discarding of
information. What if we imagine adding the restriction of no
classical communication?  This doesn't work either: the EoA is
still not a monotone. What about adding the restriction of no
generalized measurements, that is, forbidding the use of an
ancilla?  One finds that the EoA is \emph{still} not a monotone.
In fact, the central example of our paper, the state~(\ref{pure}),
shows that even if one forbids all operations save one, namely, a
single two-outcome projective measurement implemented by a single
party, the EoA is still not found to be a monotone because it can
be increased by this operation. Thus, we conclude that there is no
sense in which the EoA can be considered to be an operational
measure of entanglement for bipartite systems.




For the rest of this article we consider the EoA to be a
\emph{tripartite} measure of entanglement and we address the
question of whether or not it is a tripartite monotone, that is,
whether or not it is nonincreasing under LOCC operations
\emph{among all three parties}. To establish the failure of
monotonicity of the EoA as a tripartite measure of entanglement,
we will compare it to another tripartite measure which we shall
call the \emph{entanglement of collaboration }(EoC). It is defined
as the maximum average amount of pure-state entanglement
(quantified by the entropy of entanglement $E_{\text{Ent}}$) that
Alice and Bob can obtain starting from a tripartite state
$\left\vert \Psi \right\rangle _{ABC}$ with reduction $\rho
_{AB},$ after arbitrary \emph{tripartite} LOCC operations between
Alice, Bob and Charlie. The EoC differs from the EoA because it
allows Alice and Bob to assist Charlie in assisting them (hence
``collaboration'').

The EoA and the EoC are similarly motivated -- they both seek to
quantify how much better Alice and Bob can make use of their
entanglement given some kind of help from Charlie. It is likely
that the EoA has been studied preferentially over EoC simply
because of the greater difficulty involved in computing the EoC --
the extra level of assistance that arises in its definition
requires one to consider an arbitrary number of rounds of
communication with optimizations at every round. However, what has
perhaps gone unnoticed in opting to solve the simpler of the two
problems is that \emph{if} there exist states for which the EoC is
greater than the EoA, then the EoA is \emph{not an entanglement
monotone.} This is so because in this case it would be possible to
increase the EoA by an LOCC operation -- namely one wherein Alice
and Bob communicate with Charlie (either once or in the course of
many rounds of communication). We shall presently demonstrate an
example of a state for which this is the case.

The example makes use of a tripartite system of dimensionality
$8\times 4\times 2$. The state is:

\begin{align}
|\Phi\rangle_{ABC} =& \frac{1}{4}\Big[ (|00\rangle
+|11\rangle +|22\rangle +|33\rangle )|0\rangle   \notag \\
& + (i|00\rangle +|11\rangle -i|22\rangle -|33\rangle )|1\rangle   \notag \\
& + (|40\rangle +|51\rangle +|62\rangle +|73\rangle )|+i\rangle   \notag \\
& + (i|40\rangle +|51\rangle +i|62\rangle +|73\rangle )|-i\rangle \Big]
\label{pure}
\end{align}
where $|\pm i\rangle\equiv (|0\rangle\pm i|1\rangle)/\sqrt{2}$.

An optimal collaborative scheme proceeds as follows. Alice first
implements a measurement with Kraus operators $ |0\rangle \langle
0|+|1\rangle \langle 1|+|2\rangle \langle 2|+|3\rangle \langle 3|$
and $|4\rangle \langle 4|+|5\rangle \langle 5|+|6\rangle \langle
6|+|7\rangle \langle 7|$ and sends the outcome to Charlie. If the
first outcome has occurred, Charlie measures the $\{\left\vert
0\right\rangle ,\left\vert 1\right\rangle \}$ basis, otherwise he
measures the $\{\left\vert +i\right\rangle ,\left\vert
-i\right\rangle \}$ basis. In each case, regardless of the
outcome, he is sure to leave Alice and Bob with a state that is
maximally entangled in the $8\times 4$ Hilbert space. The scheme
is optimal because it always achieves a maximally entangled state.
When measured by the entropy of entanglement, the EoC in this case
is $2$ e-bits.

On the other hand, if no communication from Alice and Bob to
Charlie is allowed, then he can not create a state that is
maximally entangled between Alice and Bob in the $8\times 4$
Hilbert space so that the EoA is strictly less than $2$ e-bits.

The proof is as follows. First note that the state $\Psi$ above
can be written as follows:
\begin{equation}
|\Phi\rangle_{ABC}=|u_{0}\rangle_{AB}|0\rangle_{C}+|u_{1}\rangle_{AB}|1\rangle_{C}
\end{equation}
where the $8\times 4$ sub-normalized states $|u_{0}\rangle_{AB}$
and $|u_{1}\rangle_{AB}$ are given by their Schmidt decomposition forms:
\begin{align}
& |u_{0}\rangle _{AB} \equiv\sum_{k=0}^{3}|c_{0,k}\rangle|k\rangle  \equiv\frac{1}{4}\Big[
(|0\rangle+z|4\rangle)|0\rangle\nonumber\\
& \;+(|1\rangle+\sqrt{2}|5\rangle)|1\rangle
+(|2\rangle+z|6\rangle)|2\rangle+(|3\rangle+\sqrt{2}|7\rangle)|3\rangle
\Big]\nonumber\\
& |u_{1}\rangle _{AB} \equiv\sum_{k=0}^{3}|c_{1,k}\rangle|k\rangle  \equiv\frac{1}{4}\Big[
(i|0\rangle+z|4\rangle)|0\rangle+|1\rangle|1\rangle\nonumber\\
& \;\;\;\;\;\;\;\;\;\;\;\;\;\;\;\;\;\;\;\;\;\;\;\;\;
\;\;\;\;\;\;\;\;\;\;+(-i|2\rangle+z|6\rangle)|2\rangle-|3\rangle|3\rangle
\Big]
\label{u}
\end{align}
where $z\equiv(1+i)/\sqrt{2}$. The reduced density matrix $\rho
_{AB}=Tr_{C}|\Phi \rangle_{ABC} \langle \Phi |$ is therefore given
by:
\begin{equation} \rho _{AB}=|u_{0}\rangle _{AB}\langle
u_{0}|+|u_{1}\rangle _{AB}\langle u_{1}| \label{rhoab}
\end{equation}

We now show that Charlie can not always prepare a state that is
maximally entangled in $8\times 4$.  In fact, we will show
something stronger: Charlie can not even prepare an $8\times 4$
state of maximal entanglement with some probability less then one.
Given the HJW theorem, it is enough to show that no convex
decomposition of $\rho _{AB}$ contains any such states. To see
this, note that any state in some decomposition of $\rho _{AB}$ is
proportional to a linear combination of $|u_{0}\rangle _{AB}$ and
$|u_{1}\rangle _{AB}$. Therefore, it is enough to show that any
normalized state of the form $x|u_{0}\rangle _{AB}+y|u_{1}\rangle
_{AB},$ with $x $ and $y$ complex, is not an $8\times 4$ maximally
entangled state.

If $x|u_{0}\rangle _{AB}+y|u_{1}\rangle _{AB}$ is maximally entangled
then all of its Schmidt coefficients
must be of equal magnitude (and non-zero). The magnitudes of these
coefficients are: \begin{align}
\lambda _{1}& =\frac{1}{16}\left( |x+iy|^{2}+\frac{1}{2}|x+y|^{2}\right)   \notag \\
\lambda _{2}& =\frac{1}{16}\left( |x+y|^{2}+2|x|^{2}\right)   \notag \\
\lambda _{3}& =\frac{1}{16}\left( |x-iy|^{2}+\frac{1}{2}|x+y|^{2}\right)   \notag \\
\lambda _{4}& =\frac{1}{16}\left( |x-y|^{2}+2|x|^{2}\right) \;.
\end{align} Thus, $\lambda _{1}=\lambda _{2}=\lambda _{3}=\lambda
_{4}$ iff $x=y=0$.  Consequently, any normalized state
$x|u_{0}\rangle _{AB}+y|u_{1}\rangle _{AB}$ cannot be maximally
entangled. This concludes the proof.


\textbf{Entanglement of assistance for more than one copy ---} We
define the $n$-copy EoA as follows:
\begin{equation}
{E}^{(n)}_{\text{Ast}}(|\Psi\rangle _{ABC})\equiv
\frac{1}{n}E_{\text{Ast}}\left(\rho _{AB}^{\otimes n}\right)\;,
\end{equation} where $|\Psi\rangle _{ABC}$ is a tripartite pure
state and $\rho _{AB}\equiv \text{Tr}_{C}|\Psi\rangle
_{ABC}\langle\Psi|$.

It follows from the results in~\cite{Smo05,Hor05}, that in the
\emph{asymptotic} limit of infinitely many copies of a tripartite (multipartite)
\emph{pure} state, the EoA (LE) is indeed an entanglement monotone.
That is, the function 
${E}^{(\infty)}_{\text{Ast}}(|\Psi\rangle _{ABC})
\equiv\lim_{n\rightarrow\infty}{E}^{(n)}_{\text{Ast}}(|\Psi\rangle
_{ABC})$
does not increase under tripartite LOCC~\footnote{The question
remains open for \emph{mixed} tripartite states.}. However, for
any {\em finite} integer $n<\infty$, the function
${E}^{(n)}_{\text{Ast}}$ is not an entanglement monotone. This
follows from the fact that for the state $|\Phi\rangle$ given in
Eq.~(\ref{pure}), the EoC is 2 ebits (having additional copies
does not change this fact), while
${E}^{(n)}_{\text{Ast}}(|\Phi\rangle)< 2$, as we now show.

From Eq.~(\ref{rhoab}) and the HJW theorem it follows that any
$8^n\times 4^n$-dimensional state in some decomposition of $\rho
_{AB}^{\otimes n}$ can be written as a linear combination of the
$2^n$ states $|u_{i_{1}}u_{i_{2}} \cdots u_{i_{n}}\rangle$, where
$i_m\in\{0,1\}$ ($m=1,2,...,n$). Thus, we want to show that any
linear combination of $|u_{i_{1}}u_{i_{2}} \cdots
u_{i_{n}}\rangle$ is not proportional to a maximally entangled
state in the $8^n\times 4^n$-dimensional Hilbert space. We will
prove it by induction.

For $n=1$ we have proved it above. Let us now assume that any
linear combination of $|u_{i_{1}}u_{i_{2}} \cdots
u_{i_{n}}\rangle$ is not proportional to a maximally entangled
state and let $|\chi\rangle$ be a linear combination of the
$2^{n+1}$ states $|u_{i_{1}}u_{i_{2}} \cdots
u_{i_{n}}u_{i_{n+1}}\rangle$. Thus, $|\chi\rangle$ can be written
as $|\chi_0\rangle|u_{0}\rangle+|\chi_1\rangle|u_1\rangle$, where
$|\chi_0\rangle$ and $|\chi_1\rangle$ are some linear combinations of
the states $|u_{i_{1}}u_{i_{2}} \cdots u_{i_{n}}\rangle$.

Let $b\in \{0,1\}$. From Eq.~(\ref{u}) and the property that
$\langle c_{b,k} | c_{b',k'} \rangle= 0$ if $k \ne k'$, one can
deduce that $|\chi_b \rangle$ can be written in the form $|\chi_b
\rangle =\sum_{s=0}^{4^n-1} |\alpha_{b,s}\rangle |s\rangle$, where
the subnormalized states, $|\alpha_{b,s}\rangle$, satisfy $\langle
\alpha_{b,s} | \alpha_{b',s'} \rangle=0$ if $s \ne s'$. Thus, it
follows that the Schmidt decomposition of $|\chi\rangle$ has the
form $|\chi\rangle= \sum_{s=0}^{4^n-1}\sum_{k=0}^{3}
\left(\sum_{b=0}^1 |\alpha_{b,s}\rangle
|c_{b,k}\rangle\right)|s\rangle|k\rangle$. Defining
$\mu_{b,s}\equiv \langle \alpha_{b,s}|\alpha_{b,s} \rangle$ and
$\eta_{s}\equiv \langle \alpha_{0,s}|\alpha_{1,s} \rangle$, the
Schmidt coefficients of $|\chi\rangle$ may be expressed as
$\lambda_{s,k}= \sum_{b=0}^1 \mu_{b,s} \langle c_{b,k} |c_{b,k}
\rangle +2 \mathrm{Re}[\eta_{s} \langle c_{0,k} | c_{1,k}\rangle
]$. From Eq.~(\ref{u}), one can deduce that
\begin{align} \lambda_{s,0} &
= \frac{1}{8}\mu
_{0,s}+\frac{1}{8}\mu_{1,s}+\frac{1}{8}\mathrm{Re}\left[(1+i)\eta_{s}\right]
\nonumber\\
\lambda_{s,1} & = \frac{3}{16}\mu
_{0,s}+\frac{1}{16}\mu_{1,s}+\frac{1}{8}\mathrm{Re}\left[\eta_{s}\right]
\nonumber\\
\lambda_{s,2} & = \frac{1}{8}\mu
_{0,s}+\frac{1}{8}\mu_{1,s}+\frac{1}{8}\mathrm{Re}\left[(1-i)\eta_{s}\right]
\nonumber\\
\lambda_{s,3} & = \frac{3}{16}\mu
_{0,s}+\frac{1}{16}\mu_{1,s}-\frac{1}{8}\mathrm{Re}\left[\eta_{s}\right]\;.
\end{align}
For $|\chi\rangle$ to be proportional to a maximally entangled
state, the $\lambda_{s,k}$ must be equal, which implies that
$\eta_{s}=0$, $\mu _{0,s}=\mu_{1,s}$ for all $s$ and
$\mu_{b,s}=\mu_{b,s'}$ for all $s$ and $s'$. But by the assumption
of our inductive proof, $|\chi_b\rangle$ is not maximally
entangled and therefore $\mu_{b,s} \ne \mu_{b,s'}$ for some $s$
and $s'$ and so $|\chi\rangle$ is not maximally entangled either.
This concludes the inductive proof.

So we see that the n-copy EoA, $E^{(n)}_{\text{Ast}}$, is not a
monotone whereas the asymptotic version of this quantity,
${E}^{(\infty)}_{\text{Ast}}
\equiv\lim_{n\rightarrow\infty}{E}^{(n)}_{\text{Ast}}$, is. Thus,
we have found that regularization yields monotonicity.  It is
interesting to note that this also occurs for the classical
information deficit, a measure of the classical correlations of a
quantum state~\cite{Hor03,Syn04}. Specifically, the single-copy
version of the classical deficit is \emph{not} a monotone, as
shown in Ref.~\cite{Syn04}, while the results of Ref.~\cite{Dev04}
demonstrate that the regularized version of the classical deficit
is equal to the one-way distillable common
randomness~\cite{Hen01,Dev03}, which \emph{is} a
monotone~\footnote{Another example of 2-way classical communication
yielding an advantage over 1-way classical communication is the
fact that the 2-way distillable entanglement is greater than the
1-way distillable entanglement for certain mixed
states~\cite{Ben96}.}.


\textbf{Generalizations of the entanglement of assistance ---}
Although the EoA was defined in terms of the entropy of
entanglement in Eq.~(\ref{originalEoA}), one could equally well
have chosen any measure of entanglement for bipartite pure states
without compromising the appropriateness of the name
``entanglement of assistance". We therefore
follow~\cite{Lau03,Pop05,GMS,Gour05} in generalizing the
definition of EoA to constitute a family of measures, with each
one being parasitic on (i.e. defined in terms of) a different
bipartite measure of entanglement. The latter is required to be an
entanglement monotone with respect to LOCC operations on the
bipartite system. We shall call it the \emph{root entanglement
measure} and denote it by $E_{\text{Rt}}.$\

Thus, given a quantum state of a tripartite system, the EoA is
defined as the maximum average of the root entanglement measure
that Alice and Bob can share after Charlie has performed a local
operation and communicated the result to Alice and Bob. Note that
this manner of defining the EoA extends readily to \emph{mixed}
tripartite states as long as $E_{\text{Rt}}$ is a measure of
entanglement that is defined for mixed bipartite states. Similarly
we can define the EoC for any given root
entanglement measure.

The example of Eq.~(\ref{pure}) demonstrates that the EoC is
greater than the EoA as long as the root entanglement measure can
distinguish maximally entangled states from nonmaximally entangled
states, which is to say that it assigns different values to these
states.
For pure states, this is a property
that is satisfied, for instance, by the von Neumann entropy but
not, for instance, by the Schmidt number. Thus our example implies
the following general result:\emph{\ the EoA fails to be an
entanglement monotone for any root entanglement measure that can
distinguish maximally entangled states from nonmaximally entangled
states. }

Our results are also significant for a quantity known as the
\emph{localizable entanglement} (LE). Given a state of a
multipartite system, the LE is the maximum average of the root
entanglement measure that Alice and Bob can share after LOCC
operations among the other parties~\footnote{One can consider
imposing restrictions on the nature of the local operations, such
as allowing only projective measurements, in order to facilitate
the optimization, as has been done in Ref.~\cite{Pop05}, but we
shall not do so here.}. The generalization of LE wherein one
optimizes over LOCC operations
between all the parties, we call the \emph{collaboratively
localizable entanglement} (CLE). Clearly, the EoA (EoC) is simply
the LE (CLE) in the case where there is only one additional party.
As such, our tripartite example demonstrates that CLE may be
greater than the LE and consequently we conclude that the LE is
not an entanglement monotone either.

\textbf{An example of smaller dimensionality using a mixed
tripartite state ---} Here we give an example of a mixed state on
a tripartite system of dimensionality $4\times 2\times 2$ for
which the EoC is greater than the EoA. The state is an equal
mixture of the four pure states
\begin{align}
|\phi _{1}\rangle_{AB} |0\rangle _{C} & \equiv
\frac{1}{\sqrt{2}}(\left\vert 0\right\rangle _{A}\left\vert
1\right\rangle _{B}+\left\vert 1\right\rangle _{A}\left\vert
0\right\rangle _{B})\left\vert
0\right\rangle _{C}, \\
|\phi _{2}\rangle_{AB} |1\rangle_{C} & \equiv
\frac{1}{\sqrt{2}}(\left\vert 0\right\rangle _{A}\left\vert
0\right\rangle _{B}+\left\vert 1\right\rangle _{A}\left\vert
1\right\rangle _{B})\left\vert 1\right\rangle _{C}. \\
|\phi_{3}\rangle_{AB} |+\rangle_{C} & \equiv
\frac{1}{\sqrt{2}}(\left\vert 2\right\rangle _{A}\left\vert
1\right\rangle _{B}+\left\vert 3\right\rangle _{A}\left\vert
0\right\rangle _{B})\left\vert +\right\rangle _{C}, \\
|\phi _{4}\rangle_{AB} |-\rangle _{C} & \equiv
\frac{1}{\sqrt{2}}(\left\vert 2\right\rangle _{A}\left\vert
0\right\rangle _{B}+\left\vert 3\right\rangle _{A}\left\vert
1\right\rangle _{B})\left\vert -\right\rangle _{C},
\end{align}
In the optimal collaborative scheme, Alice implements a projective
measurement with Kraus operators $\left\vert 0\right\rangle
\left\langle 0\right\vert +\left\vert 1\right\rangle \left\langle
1\right\vert$ and $\left\vert 2\right\rangle \left\langle
2\right\vert +\left\vert 3\right\rangle \left\langle 3\right\vert
$ and communicates her outcome to Charlie, who then knows whether
to measure in the $\{\left\vert 0\right\rangle ,\left\vert
1\right\rangle \}$ or $\{\left\vert +\right\rangle ,\left\vert
-\right\rangle \}$ bases in order to be sure to collapse $AB$ to
one of the $\left\vert \phi _{i}\right\rangle .$  The scheme is
optimal because these are maximally entangled in the $4\times 2$
space.  A scheme consisting purely of assistance by Charlie cannot
do as well because for any generalized measurement that Charlie
performs, he leaves $AB$ in a mixture of the $|\phi_i\rangle$ with
at least three of these states receiving non-zero weight, and it
can be shown that no such mixture is maximally entangled in
$4\times 2$.

\textbf{Discussion ---} In the presence of a noisy environment,
creating and distributing entanglement on a multi-user quantum
network is a challenging problem.  The question of precisely how
much entanglement can be created between a chosen pair of parties
starting from an arbitrary multi-party state is therefore an
important one. In the introduction it was argued that any measure
of entanglement that sought to quantify the degree of success of
some distributed QIP task must be a monotone under LOCC, and
entanglement creation on a network certainly qualifies as such a
task. This leads us to the following puzzle. The EoA and LE
certainly \emph{seem} to have some operational significance for
the problem of entanglement creation, and yet we have just shown
that they fail to be monotones. How can this be?

The puzzle is resolved by noting that the EoA and LE \emph{do}
quantify the usefulness of a quantum state for certain tasks, but
these are not the sorts of tasks that one typically considers,
namely, distributed QIP tasks given LOCC. Rather, they quantify
the usefulness of a quantum state for distributed QIP tasks given
local operations and \emph{restricted} classical communication,
where the restriction is that there is \emph{no classical
communication allowed from Alice and Bob to the other parties}.
The EoA and LE are strictly non-increasing under these restricted
operations. We may say that they are monotones, but only with
respect to this restricted version of LOCC.

It follows that EoA and LE are operationally significant
\emph{only if} one is faced with a network that places the
specified restrictions on classical communications. However, such
a restriction is unlikely to arise naturally in practical quantum
information networks and consequently for such applications it is
the EoC and CLE rather than the EoA and LE that are the quantities
that are of the most interest.

In addition to its obvious applications in the field of quantum
communication and the study of multipartite entanglement, the
concept of LE has proven to be fruitful in the field of condensed
matter physics, in particular for understanding the complex
physics of strongly correlated states (for example, see references
in~\cite{Pop05}). Our results are therefore likely to have
applications in this area. For instance, the \emph{entanglement
length}, which characterizes the typical length up to which
bipartite entanglement can be localized in the system, can be, in
general, greater when defined in terms of the CLE rather then the LE.

There was a significant caveat in the statement of our main
result: we have only shown that EoA and LE fail to be entanglement
monotones (the usual kind, with respect to unrestricted LOCC) when
they are defined in terms of a root entanglement measure that can
distinguish maximally entangled states from nonmaximally entangled
states. This leaves open the possibility that the EoA and LE can
be entanglement monotones for choices of the root measure that are
not of this sort. Elsewhere we show that this possibility is
indeed realized in the case where the root measure is the
G-concurrence, a generalization of the concurrence~\cite{Woo98} to
pairs of systems of dimensionality greater
than~2~\cite{Gou05,Gour05}. Also, if Alice and Bob's systems are
of smaller dimensionality than considered here, then there will
likely be \emph{more} root entanglement measures for which the EoA
is found to be a monotone. Indeed, for $2\times2\times2$ pure
states if the root measure is the concurrence (which \emph{does}
distinguish maximally from nonmaximally entangled states), then
the resulting EoA can be proven to be a monotone \cite{GMS}.
Determining in which Hilbert spaces and for which root
entanglement measures the EoA is a monotone will help to identify
those distributed QIP tasks for which having a collaboration with
Charlie provides no advantage over merely having his assistance.

Acknowledgments:--- The authors gratefully acknowledge Michal
Horodecki, Debbie Leung, Michael Nielsen, Jonathan Oppenheim and
Martin Plenio for helpful comments. G.G.\ acknowledge support by the
National Science Foundation (NSF) under Grant No.\ ECS-0202087.

\end{document}